\documentclass[aps,prb,reprint,showpacs,showkeys,superscriptaddress,preprintnumbers,amsmath,amssymb,longbibliography]{revtex4-2}
\usepackage{graphicx}
\usepackage{dcolumn}
\usepackage{amsmath} 
\usepackage{amssymb}
\usepackage{color}
\usepackage{soul}
\usepackage{xcolor}
\usepackage{multirow}
\usepackage{chemfig}

\usepackage{bm}

\begin{document}

\title{Emergent Random Spin Singlets in Disordered Spin-1/2 perovskite BaCu$_{1/3}$Ta$_{2/3}$O$_3$}

\author{Sagar Mahapatra}
\affiliation{Department of Physics, Indian Institute of Science Education and Research, Pune 411008, Maharashtra, India}

\author{Francesco De Angelis}
\affiliation{Dipartimento di Scienze, Universitá Roma Tre, I-00146 Roma, Italy}

\author{Dibyata Rout}
\affiliation{Department of Physics, Indian Institute of Science Education and Research, Pune 411008, Maharashtra, India}

\author{Priyanshi Tiwari}
\affiliation{UGC-DAE Consortium for Scientific Research, University Campus, Khandwa Road, Indore 452 001, India}

\author{Martin Etter}
\affiliation{Deutsches Elektronen-Synchrotron (DESY), Notkestra\ss e 85, 22607 Hamburg, Germany}

\author{Edmund Welter}
\affiliation{Deutsches Elektronen-Synchrotron (DESY), Notkestra\ss e 85, 22607 Hamburg, Germany}

\author{M. P. Saravanan}
\affiliation{UGC-DAE Consortium for Scientific Research, University Campus, Khandwa Road, Indore 452 001, India}

\author{Rajeev Rawat}
\affiliation{UGC-DAE Consortium for Scientific Research, University Campus, Khandwa Road, Indore 452 001, India}

\author{Satoshi Nishimoto}
\affiliation{IFW Dresden, Helmholtzstr. 20, 01069 Dresden, Germany}

\author{Carlo Meneghini}
\affiliation{Dipartimento di Scienze, Universitá Roma Tre, I-00146 Roma, Italy}

\author{Surjeet Singh}
\email{surjeet.singh@iiserpune.ac.in}
\affiliation{Department of Physics, Indian Institute of Science Education and Research, Pune 411008, Maharashtra, India}
\affiliation{IFW Dresden, Helmholtzstr. 20, 01069 Dresden, Germany}

\begin{abstract}
We investigate the disordered perovskite BaCu$_{1/3}$Ta$_{2/3}$O$_3$, where Cu (spin-1/2) and Ta randomly occupy a pseudo-cubic lattice. Synchrotron X-ray diffraction and X-ray absorption spectroscopy establish the local nature of the disorder, revealing the presence of structurally constrained magnetic exchange paths. No magnetic ordering or spin freezing is observed down to 0.1 K. The low-temperature magnetic and thermodynamic behavior is captured by a broad but non-singular distribution $P(J)$ of exchange couplings $J$. These results open the possibility of realizing a disordered quantum ground state where the exchange randomness is broad yet intrinsically bounded, departing from the conventional infinite-randomness fixed point driven random-single phase.
\end{abstract}

\maketitle

Quenched disorder in quantum spin systems has been a topic of considerable interest. In one-dimensional Heisenberg antiferromagnetic spin-1/2 model, the quenched disorder results in a novel Random Singlet (RS) ground state, arising from an infinite-randomness fixed point in the real-space renormalization-group flow. The RS phase comprises of spin dimers at arbitrarily large separations, with the exchange couplings $J$ described by a broad, scale-free probability distribution $P(J)$~\cite{Dasgupta_PhysRevB.22.1305, ma1979, Fisher1994}.
 
Importantly, the RS phase has also been shown to emerge in highly frustrated quantum magnets with quenched disorder~\cite{kimchi_2018}. In the absence of disorder, these systems exhibit novel quantum spin liquid ground states. However, when disorder is included, the RS phase emerge as a competing ground state. The effective coupling between the spins forming the RS phase follows a scale-free, power-law distribution $P(J) \sim J^{-\gamma}$, leading to thermodynamic quantities diverging at low-temperatures as: $\chi \propto T^{-\gamma}$, $c\rm_{mag} \propto T^{1-\gamma}$, where $\chi$ and $c\rm_{mag}$ are magnetic susceptibility and magnetic specific heat, respectively.

An important  consequence of effective $J^{-\gamma}$ distribution is the emergence of scale-invariant behavior, manifested by the collapse of $M T^{\gamma-1}$ versus $\mu_0 H/T$, and $(\mu_0 H)^\gamma c\rm_{mag}/T$ versus $T/\mu_0 H$ plots over a wide range of temperatures and magnetic fields (henceforth referred to as $M[H, T]$- and $c[H, T]$-scaling, respectively). Such scaling behavior has been demonstrated for various frustrated spin systems with quenched disorder, including LiZn$_2$Mo$_3$O$_8$~\cite{LiZn2012}, ZnCu$_3$(OH)$_6$Cl$_2$~\cite{22_kimchi_nat_com_scaling}, YbMgGaO\(_4\) \cite{YbMg_2017}, YbZnGaO\(_4\) \cite{YbZn2018}, Y\(_2\)CuTiO\(_6\) \cite{18_SKundu_YCTO}, Sr$_2$IrO$_{6-\delta}$ \cite{SrIr2O6}, and Ba\(_2\)YMoO\(_6\) \cite{BaY2010}.

Recently, the RS phase has been proposed for the perovskite spin-1/2 cuprates, SrCu$_{1/3}$\textit{M}$_{2/3}$O$_3$ (\textit{M} = Ta or Nb)~\cite{hossain2024evidence,sana2024possible}. In these perovskites the B-site forms an pseudo-cubic lattice ($c/a~\approx~1$), randomly occupied by magnetic Cu and non-magnetic Ta(Nb). These compounds do not exhibit magnetic long-range ordering or spin freezing down to the lowest temperatures. Their low-temperature magnetic and thermodynamic properties have been argued to be compliant with the the RS ground state. 

The 33\% spin-1/2 occupancy (Cu/M $\equiv 1/2$), which exceeds the 31\% percolation threshold of a cubic lattice, raises question about the apparent absence of large Cu-\textcolor{black}{connected} clusters that  typically favor tendency towards  magnetic ordering or spin freezing 
phase~\cite{Coles01041978}. This naturally raises question about the precise nature of Cu-\textcolor{black}{\textit{M}} disorder in these perovskites? In particular, what makes this disorder sufficiently strong to generate $P(J)\propto J^{-\gamma}$ \textcolor{black}{distribution? } Here, we attempt to understand these questions and, in conjunction, we also closely examine the nature of the ground state in these materials: i.e., given the non-frustrated lattice and high concentration of magnetic spins, what makes the low-energy effective distribution vary as $J^{-\gamma}$?

To address these questions, here we examine BaCu$_{1/3}$Ta$_{2/3}$O$_3$ (BCTO). True to its sister compounds, no long- or short-ranged magnetic ordering is detected in BCTO down to the lowest measured temperature of 0.1 K. The low-temperatures specific heat is characterized by a broad Schottky anomaly with associated entropy saturating around $0.4R\ln2$, indicating that a large fraction of the spins remain dynamic down to 0.1 K. $\chi(T)$ exhibits a power-law-like divergence at low temperatures for $\gamma \approx 0.67$. The M(H) plots in the same temperature range show M[H, T]-scaling. However, the $c[H, T]$-scaling or c$_{mag} \propto T^{1-\gamma}$~are not observed 
for any non-zero $\gamma$ (specifically $\gamma = 0.67$), suggesting that the magnetic ground state is more complex than the RS phase.

To gain understanding of the BCTO atomic structure, we used synchrotron-based powder X-ray diffraction (XRD) and X-ray absorption spectroscopy (XAS) to probe the long range (crystallographic) and local atomic structure, respectively. We show that Cu and Ta  distribution is statistically homogeneous throughout the lattice, without evidence for macroscopic phase separation. However, the local structure reveals a prevalence of hetero-atomic Cu-O-Ta pairs with roughly equal numbers of Cu and Ta around the Ta absorber (Ta-Cu/Ta \(\approx\) 2.9/3.1), but a Ta rich environment around the Cu absorber (Cu-Cu/Ta \(\approx\) 0.4/5.6), leading to magnetic exchange paths such as Cu-O-Ta-O-Cu, Cu-O-Ta-O-Ta-O-Cu, and so on. The EXAFS results ensures the occasional presence of a Cu-O-Cu type exchange, while the roughly 50\% probability of finding Cu or Ta around Ta ensures that the average Cu–Cu separation does not grow arbitrarily, thuls preventing $P(J)$ from becoming singular as J \(\rightarrow 0 \). 
 
We thus analyzed the experimental data using a data-driven distributed exchange dimer model, which yields a very broad $P(J)$; however, unlike a scale-invariant $J^{-\gamma}$ form, the derived $P(J)$ is non-singular, featuring a broad peak centered around 4 K with additional shoulder-like features near 70 K and 0.1 K. This form of $P(J)$, characterized by a long but non-diverging low-$J$ tail, is consistent with the Cu-Ta disorder revealed by EXAFS, allowing a unified description of the temperature and field dependence of susceptibility and specific heat. 

\begin{figure}[htp]
    \centering
   \includegraphics[width=\linewidth]{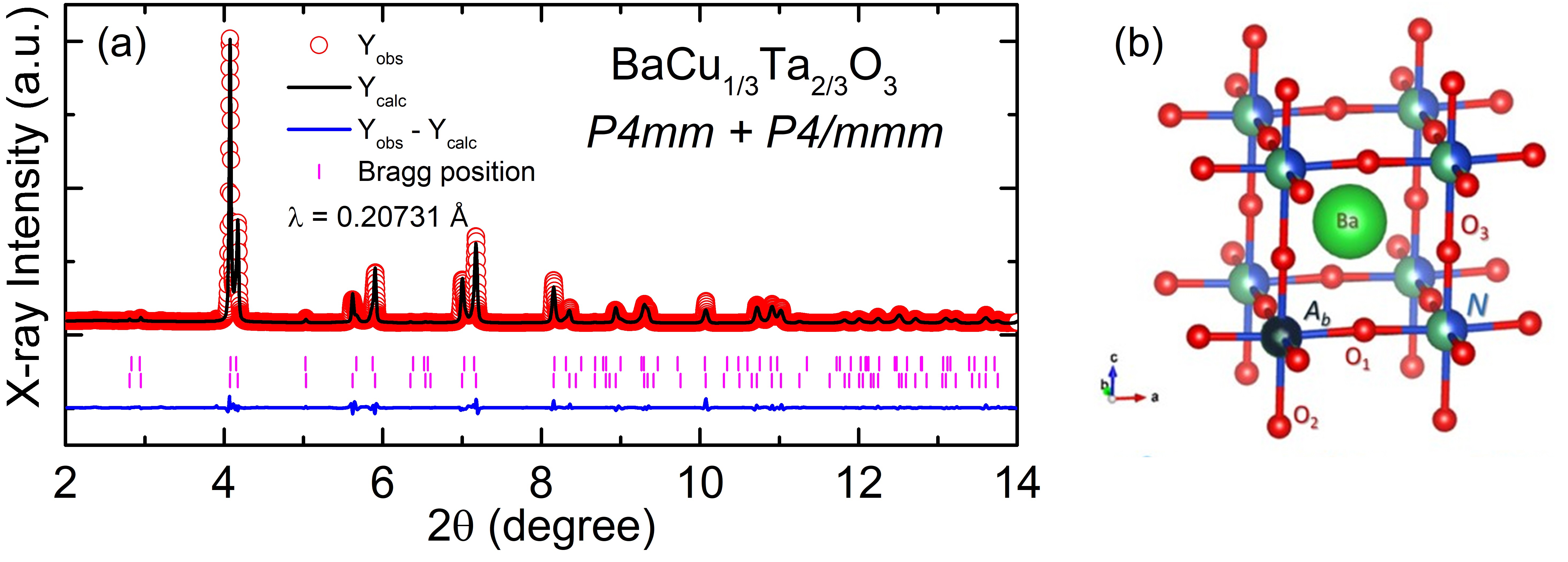}
    \caption{(a) The Rietveld refinement of X-ray powder diffraction patterns of BCTO, (b) a schematic of the unit cell of the BCTO crystal structure. Here, A$_b$ and N are the absorbing atom and the nearest-neighbour atom, respectively, as discussed in the XAS (see text).}
    \label{fig:xrd}
\end{figure}

The BCTO sample was synthesized by a conventional solid-state reaction at 1100$^\circ$C. It adopts an ABO$_3$-type tetragonal perovskite structure, comprising a pseudo-cubic unit cell whose center is occupied by A = Ba and corners randomly by B~=~Cu and Ta in $1:2$ ratio, with Cu/Ta–O–Cu/Ta bonds align along the edges [Fig.~\ref{fig:xrd}(b)]. Detailed synchrotron XRD analysis reveals two closely related tetragonal phases, \textit{P4mm} and \textit{P4/mmm}. 
The main crystallographic difference between the two structures is a very slight off-centering of the B-site in the \textit{P4mm} phase, which primarily contributes to further broadening of $J$, as will become clear in the subsequent discussion. For details of structural refinement, refer to the Supplementary Material~\cite{SM}.
\\

\begin{figure}[b]
    \centering
         \includegraphics[width=0.44\linewidth]{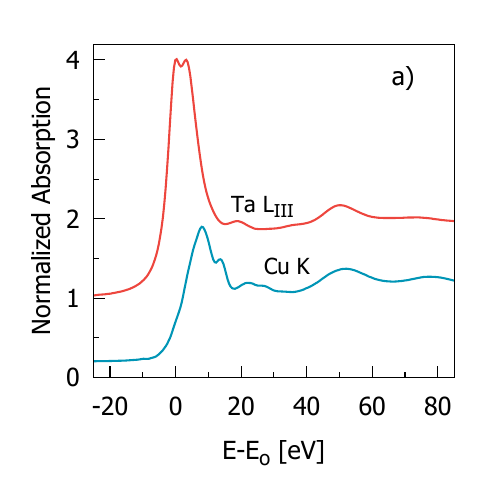}    
    \includegraphics[width=0.44\linewidth]{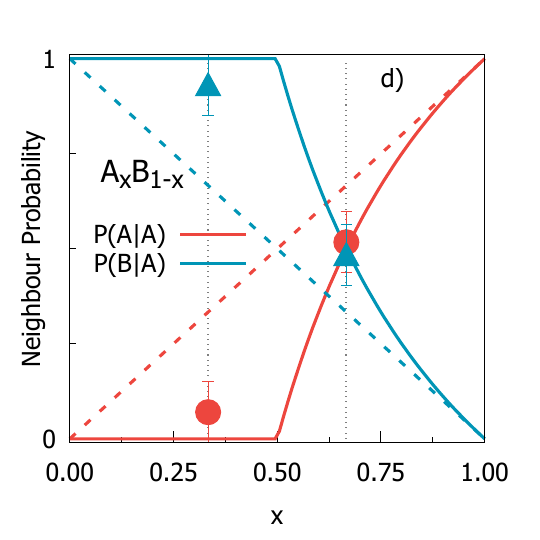}
    \includegraphics[width=0.88\linewidth]{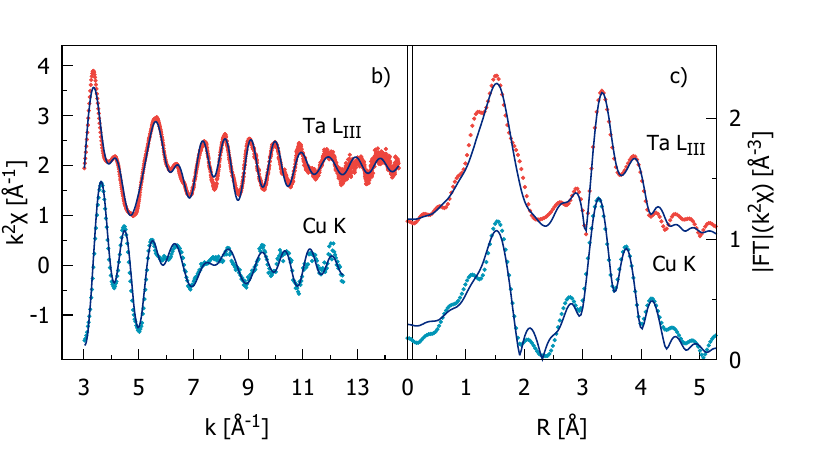}

    \caption{ \textcolor{black}{(a) Normalized Ta-L\textsubscript{III} and Cu-K edge XANES spectra, with the energy scale given relative to the edge energy (E–E\textsubscript{o}) to facilitate comparison; (b) \(k^2\)-weighted experimental EXAFS data (dots) and corresponding best-fit curves (solid lines) at the Cu and Ta edges for the BCTO sample; (c) moduli of the \(k^2\)-weighted \(\vert \rm{FT}\vert\) of the experimental data (dots) and best fit curves (solid lines) for the various spectra; the curves are vertically offset for clarity. (d) homo-atomic (red) and hetero-atomic (blue) neighbour probabilities around the absorber plotted as a function of absorber concentration \textit{x}, for both random distribution (dashed lines) and chemical order (solid lines) models (see text), compared with the experimental data, where \textit{x} = 0.33 (0.66) corresponds to the Cu (Ta) edge results.}}
    \label{fig:EXAFS_fit}
\end{figure}

EXAFS results are summarized in Fig.~\ref{fig:EXAFS_fit}. The normalized Cu-K and Ta-L\textsubscript{III} edge spectra in the XANES region (Fig.~\ref{fig:EXAFS_fit}) confirm Ta$^{5+}$ and Cu$^{2+}$ charge states. The Ta-L$_{III}$ edge exhibits an intense white-line ascribed to dipole allowed photoelectron excitations from 2$p$ core to empty 5$d$ states. The bimodal shape of the white-line can be attributed to the $e_g-t_{2g}$ splitting of about 4~eV in agreement with O\textsubscript{h} coordination symmetry.~\cite{XANES_NbTa} The tiny and broad pre-edge shoulder of Cu-K suggests Jahn-Teller distorted octahedral geometry. The $k^2$-weighted experimental EXAFS spectra, \(k^2  \chi^{\text{exp}}\), and the corresponding best-fit curves at the Cu, and Ta edges are shown in Fig.~\ref{fig:EXAFS_fit}, together with the moduli of the Fourier transforms $\vert \rm{FT}\vert$ of both  $k^2$-weighted experimental data and best-fit curves. The $\vert \rm{FT}\vert$ representation provides a more intuitive view of the coordination shells around the absorber, despite distortions introduced by phase-shift effects, amplitude modulation, and interference between different scattering contributions~\cite{meneghini2012estra}.  The main peak around 1.6 \AA\ in Fig.~\ref{fig:EXAFS_fit} (squeezed by phase shift effect) is ascribed to the oxygen coordination shell around Cu or Ta absorbers. The complex features between 2.5 \AA\ and 4.5 \AA\ account for contributions coming from single-scattering (SS), and multiple-scattering (MS) paths necessitated by the nearly collinear geometry of Cu-O-Ta bonds~(see Fig.~\ref{fig:xrd}). The main results are summarized in Table~\ref{table:XAFS}. For details of EXAFS analysis, see Supplementary Material~\cite{SM}.

\begin{table}[htb]
\caption{Results of the EXAFS data refinement at the Cu K and Ta L\textsubscript{III} edges, compared with crystallographic distances based on XRD Rietveld analysis. (See Supplementary Material for details~\cite{SM}.}
\label{table:XAFS}
    \centering    
\renewcommand{\arraystretch}{1.2}
    \begin{tabular}{llllllll}

              &             & \multicolumn{3}{c}{ {Cu K-Edge}} & \multicolumn{3}{c}{ {Ta L$_{III}$-Edge}} \\
        \hline
       & R$\rm_{XRD}$      & N & R     & \(\sigma^2\)                             &N & R    & \(\sigma^2\)           \\
       & (\AA)  &   & (\AA) & \(\times 10^2\) (\AA\textsuperscript{2}) &  & (\AA)& \(\times 10^2\) (\AA\textsuperscript{2}) \\
        \hline
        O$_1$  & 2.022     &4    & 2.03(1)    & 0.87(4)                &6     &  1.98(1)   & 0.89(3) \\
        O$_2$  & 2.095     &2    & 2.32(1)    & 0.87                   &      &            &         \\
        Ba     & 3.545     &8    & 3.55(2)    & 1.3(2)                 &8     & 3.55(1)    & 0.95(8) \\
        Ta(SS) & 4.1       &5.6(3) & 4.05       & 0.67&3.1(3)& 3.93(3)    & 1.9(3)  \\

        Cu(SS) & 4.1       &0.4    & 4.19(2)    & 2.2                  &2.9   & 4.05(3)    & 0.67(5)\\

        \hline     
    \end{tabular}
\end{table}

The EXAFS analysis (Table~\ref{table:XAFS}) shows that the local coordination geometries of the CuO$_6$ and TaO$_6$ octahedra deviate significantly from the average crystallographic structure one obtains using XRD~\cite{Boyce1989}. In particular, the CuO$_6$ octahedra appear strongly Jahn–Teller distorted, with four shorter Cu–O$_1$ bonds ($\rm R_{CuO_1} \simeq 2.03$ \AA) and two longer Cu–O$_2$ bonds ($\rm R_{CuO_2} \simeq 2.30$ \AA). In contrast, a single oxygen coordination shell is found around Ta ($\rm R_{TaO} \simeq 1.98$ \AA). The averagely longer Cu–O distances are consistent with the larger ionic radius of Cu$^{2+}$ (0.73 \AA) compared with Ta$^{5+}$ (0.64 \AA). 
More importantly, EXAFS analysis makes it possible to obtain information on the chemical ordering in a binary alloy~\cite{meneghini1994structure,MeneghiniPRL2009}, as described in detail in the Appendix section.  
As depicted in Table~\ref{table:XAFS}, each Ta-absorber has roughly equal number of Ta and Cu atoms around it (N\textsubscript{Ta-Ta/Cu~}$\approx$~3.1/2.9). In contrast, each Cu absorber is surrounded by significantly more Ta than Cu (N\textsubscript{Cu-Cu/Ta}$\approx$~0.4/5.6). This finding demonstrate a tendency for chemical order in the neighbour distribution that favors the hetero-atomic correlation as opposed to to a random distribution, as evident in the Fig.\ref{fig:EXAFS_fit}. This result indicates a reduced number of Cu–O–Cu dimers (\chemfig{Cu::Cu}) in the structure compared with dimers where one  Ta atoms lie between the Cu atoms, i.e., \chemfig{Cu::Cu}~$\equiv$~Cu–O–Ta–O–Cu. 
Noticeably, because the likelihood of finding either Cu or Ta around Ta is approximately 50\%, the probability of forming Cu dimers with more than one intervening Ta atom declines rapidly.

\begin{figure}[t]
   \centering
   \includegraphics[width= 1.0\linewidth]{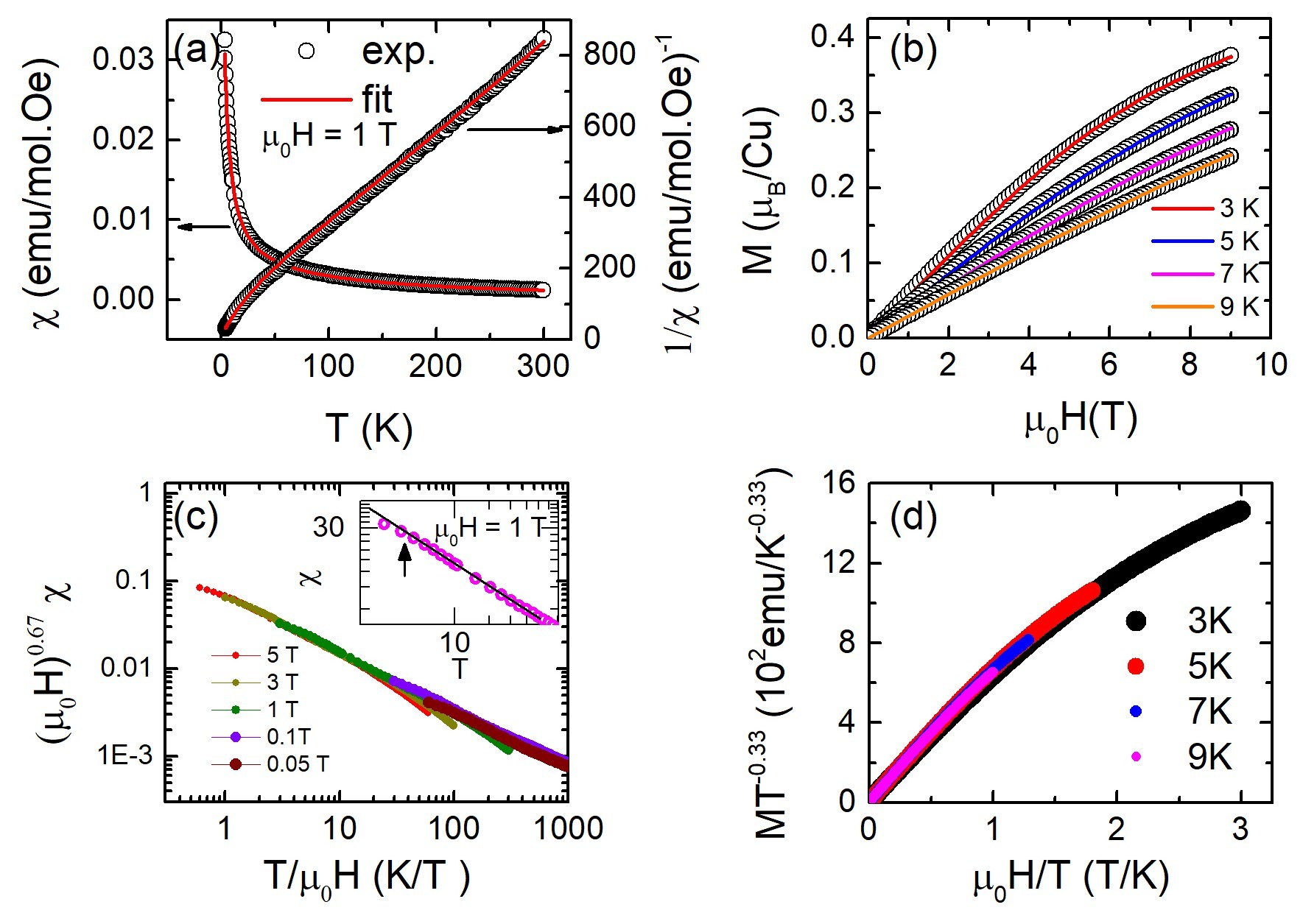}
  \caption{(a) Magnetic susceptibility $\chi$ and $\chi^{-1}$ plotted as a function of temperature; (b) Isothermal magnetization $M(H)$ at various temperatures. The solid lines in (a) and (b) are fits to the data (see text); (c) and (d) show data scaling behavior for $\chi$(T) and $M(H)$, respectively. Inset in (c) shows $\chi(T)$ on a double-log plot. The units are same as in (a). The arrow indicates departure from T$^{-\gamma}$ at low temperatures. } 
  \label{fig:BCTO_merge}
\end{figure}

We now relate these findings to the observed magnetic behvior of BCTO. The temperature dependence of magnetic susceptibility ($\chi$) is shown in Fig.~\ref{fig:BCTO_merge}(a). No signs of magnetic ordering could be seen at any temperature above 3 K (the lowest temperature in our experiment). The zero-field-cooled (ZFC) and field-cooled (FC) plots overlap indicating the absence of any spin-glass-like phase due to randomness. The Curie-Weiss fit gives an effective moment of $\rm \mu_{eff}=1.85 \pm 0.05~\mu_B$, in good accord with Cu$^{2+}$ (spin 1/2), and $\rm \theta_{CW} = -35~\pm~10$ K. As shown in Fig.~\ref{fig:BCTO_merge}(c)[inset]  $\chi(T)$ exhibits an approximate $T^{-0.67}$ divergence down to about 4 K, below which a slight deviation is seen. The main panel shows an approximate data collapse at low temperatures when $\rm (\mu_0H)^{0.67}\chi$ is plotted against $T/\mu_0H$. Similarly, M(H) plots at various temperatures between 3 K and 9 K, shown in Fig.~\ref{fig:BCTO_merge}(b), exhibit the $M[H, T]$-scaling, as shown in Fig.~\ref{fig:BCTO_merge}(d).

\begin{figure}[!]
	\centering
    \includegraphics[width=\linewidth]{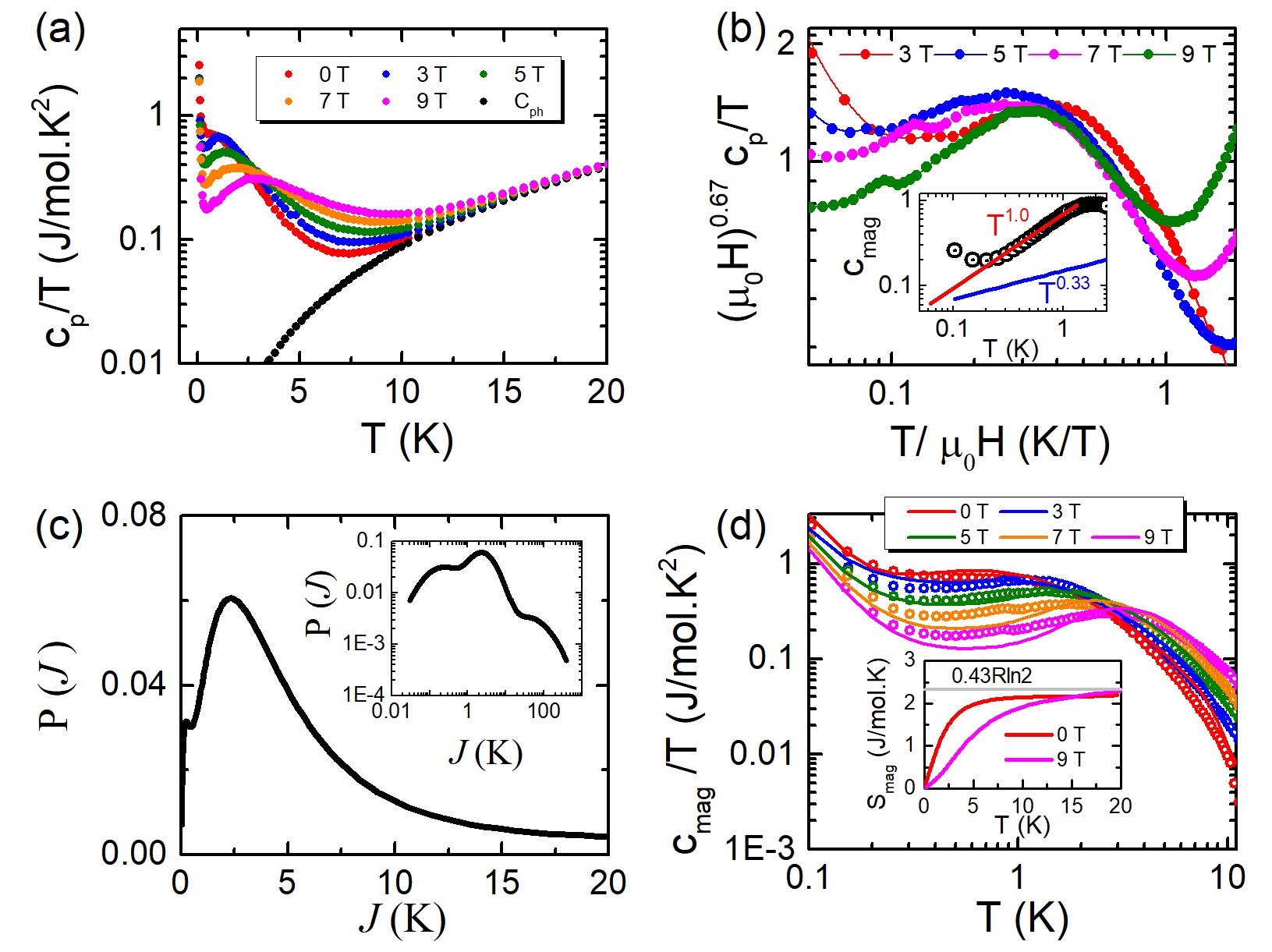}
	\caption{ (a) The temperature variation of specific heat plotted as $\rm c_p/T$ (phonon contribution is also shown); (b) $(\mu_0 H)^{\gamma} c_p/T$ ($\gamma = 0.67$) plots against T/$\mu_0H$ for various applied fields are shown not to obey the scaling behavior. Inset in (b) shows $c\rm_{mag}$ varying linearly with T (red line). T$^{0.33}$ (blue line) is also shown to emphasize non-compliance with the randoms-singlet ground state; (c) exchange-coupling distribution $P(J)$ obtained here (inset: double-logarithmic plot of the same); (d) $c_{\rm mag}(T)/T$ versus $T$ plots for various H. The solid lines in (d) are fit to the data (inset shows the temperature variation of $\rm S_{mag}$ for $\mu_0H = 0, 9$~T).}
	\label{fig:PJ}
\end{figure}

The specific heat, ploted as c$_p$/T for $\mu_0$H = \{0, 3, 5, 7, and 9 T\}, is shown in Fig,~\ref{fig:PJ}(a). It exhibits a Schottky peak at low temperatures: the peak broadens and shifts to higher temperatures with increasing $H$. The very low temperature upturn is attributed to the onset of nuclear Schottky contribution. The $\rm c_{mag}/T$ is plotted in Fig.~\ref{fig:PJ}(d). The magnetic entropy $\Delta$S$\rm_{mag}$, estimated by integrating c$\rm_{mag}$/T above 0.1 K, is shown in Fig.~\ref{fig:PJ}(d) (inset). $\rm c_{mag}$ is obtained by subtracting the phonon part (c$\rm_{ph}$) from c$_p$~(see Ref.~\cite{SM}). $\rm \Delta S_{mag}$ saturates at a value of $\approx~0.4\ln R$ near 20 K.  
This suggests a significant zero-point entropy with almost as many as 60\% of the Cu spins remaining dynamic down to temperatures as low as 0.1 K.

In the random-singlet phase c$\rm_{mag}$ is expected to diverge as $T^{1-\gamma}$ ($\gamma = 0.67$ in accordance with the $\chi$ and $M[H, T]$-scaling). However, the c$\rm_{mag}$ data below 1 K follows a T-linear behavior, as shown in the inset of Fig.~\ref{fig:PJ}(b), which deviates significantly from the expected T$^{0.33}$ form. Furthermore, 
$\rm (\mu_0H)^\gamma c_{p}/T$ versus $\rm T/\mu_0H$ plots do not show the expected data collapse for $\gamma = 0.67$ (or any other $\gamma$ value~Ref.~\cite{SM}, as shown in Fig.~\ref{fig:PJ}(b). 
There is an approximate convergence in the narrow range: $0.3 < T/\mu_0H < 1$ or $0.9~K < T < 3~K$, but the low temperature deviation is significant.
 
 In frustrated spin systems with quenched disorder, while the frustration prevents long-range ordering, the quenched disorder causes a small fraction of spins, positioned randomly due to disorder, to form the RS ground state, with the distribution of $J$'s often given by $P(J)\propto J^{-\gamma}$. As a result, the thermodynamics properties at low temperatures are dominated by these weakly coupled dimers, resulting in power-law and data-scaling behaviors. On the other hand, in BCTO with high-density of spin-1/2's placed randomly on a pseudo-cubic lattice, the expected behavior is not seen at very low temperatures.

We thus develop a unified distributed-exchange dimer model, which makes no \emph{a priori} assumption on the form of $P(J)$. Instead, $P(J)$ is reconstructed from the experimental data as explained here. For numerical flexibility, we represent $P(J)$ as a sum of log-normal components:
\begin{equation}
		\rm P(J)=\sum_{k=1}^{K} w_k
		\frac{1}{J\sigma_k\sqrt{2\pi}}
		\exp\!\left[-\frac{(\ln J-\ln J_{0k})^2}{2\sigma_k^2}\right],
		\qquad \\ \sum_k w_k=1.
		\label{eq:PJ}
\end{equation}
Here, $J_{0k}$ and $\rm \sigma_k$ are the median and width of each component, and $\rm w_k$ its normalized weight. This flexible yet compact form allows $P(J)$ to capture both narrow and broad features, enabling a direct reconstruction from the data without imposing a theoretical bias. Further details are provided in the Supplementary Material~\cite{SM}.

The model includes: (i) antiferromagnetic spin-1/2 dimers obeying $P(J)$ in eq.~\ref{eq:PJ}, and (ii) a small fraction, $f$, of paramagnetic monomers (orphan spins). 
The  expressions for $\chi(T)$, $M(H, T)$, and c$_{mag}$(H, T) are given in the Supplementary Material~\cite{SM}. All datasets, $\rm c_{\rm mag}(T, H)$, 
$\rm \chi(T)$, and $M(H,T)$, were fitted simultaneously using a global least-squares objective. The best-fit solution reproduces all datasets with an average normalized root-mean-square deviation of approximately 3\%, as shown in figure~\ref{fig:BCTO_merge}(a), (b), and~\ref{fig:PJ}(d), respectively. The monomer fraction obtained is $f\simeq0.04$.

The reconstructed $P(J)$, shown in Fig.~\ref{fig:PJ}, is very broad, spanning over two decades in energy, and characterized by a broad peak at $J \approx 4$~K with shoulder-like features at 70 K and 0.1 K, and with a long but non-diverging low-$J$ tail. As discussed before, the Ta-rich local environment around Cu suppresses direct Cu-O-Cu type dimers compared to the dimers formed with at one intervening Ta.
Similarly, the near-equal Cu/Ta coordination around Ta, sharply reduces the number of intervening Ta. Accordingly, we ascribe the high-temperature feature near 70 K to the Cu-O-Cu superexchange. This is consistent with the Weiss temperature $\theta_{CW}$, and the value of $J$ for related perovskite cuprates~\cite{7_BCSO_Zhou}). The observed decrease by an order of magnitude with an additional Ta atom in the exchange path (4 K peak) is also reasonable, as is the further reduction by another order of magnitude with two or more intervening Ta atoms (0.1 K feature and low-$J$ tail). That the number density of dimers with one intervening Ta is significantly larger is also in qualitative agreement with the distribution of Cu and Ta. 

The additional broadening of these features arises from the strong dependence of effective exchange couplings on cluster geometry, as confirmed by Exact-Diagonalization calculations (see Supplementary Material~\cite{SM}). Further, the primary $J$ for $P4/mmm$ and $P4mm$ differ slightly, the B-O-B bond angle being 180$^\circ$ and 169$^\circ$, respectively. 

Incidentally, above the $J/k_B$ = 4 K, the $P(J)$ versus $J$ plot has an approximate power-law behavior. This becomes more evident in the log-log plot shown in the inset of Fig.~\ref{fig:PJ}, which reveals an approximate linear regime above 4 K, inclined at a slope of  $\approx -0.7$. Therefore, on the high-temperature side of the 4 K peak, $P(J)$ resembles a typical power-law dependence, which may account for the apparent power-law and scaling behavior seen in our data above 3~K, but not at the low temperatures.

To summarize, we studied the disordered perovskite BCTO, where Cu and Ta randomly occupy a pseudo-cubic B-sublattice. The nature of this disorder was characterized using EXAFS. We find no evidence for long-range magnetic order or spin freezing down to 0.1 K. The magnetic ground state is best described as a collection of singlets with a broad but non-singular distribution $P(J)$ of exchange couplings, consistent with the structurally constrained exchange paths inferred from EXAFS data. Our results point to a class of three-dimensional spin-disordered materials in which disorder leads to broad, but finite, exchange distributions and does not drive the system toward an infinite-randomness fixed point.\\ 

\begin{acknowledgments}
Portions of this research were carried out at the light source PETRA-III of DESY, a member of the Helmholtz Association (HGF). We would like to thank the beamline scientists for assistance at the beamlines P65 and P02.1. Financial support by the Department of Science \& Technology (Government of India) provided within the framework of the India@DESY collaboration is gratefully acknowledged. SM would like to acknowledge the University Grants Commission (UGC), India, for financial support in the form of a research fellowship. SM is also
grateful to the I-HUB Quantum Technology Foundation (QTF) at IISER, Pune, for financial support under the Senior Research Fellowship scheme. SM acknowledges Dr V.S. Patankar Dissertation Fellowship for supporting through a one-time financial aid.

\end{acknowledgments}

\begin{center}
	{\textbf{{APPENDIX}}}
\end{center}

\section{EXAFS and chemical ordering}

To understand how EXAFS analysis can deduce the local chemical order in binary alloys,  that is the relative arrangement of neighboring species, let us consider a solid solution made of two atomic species, $A$ and $B$, randomly occupying the sites of a cubic lattice with composition $A_xB_{1-x}$. Different situations may arise. For a perfectly random distribution (R), the probability of finding a pair of neighboring atoms is determined solely by the relative concentrations of the two species. Thus, the probabilities of finding $AA$, $BB$, and $AB$ pairs are, respectively,:
\[
P_R(AA)=x^2, \; P_R(BB)=(1-x)^2, \; P_R(AB)=2x(1-x).
 \]
Alternatively, the system may exhibit a preference for either \textit{hetero-atomic} or \textit{homo-atomic} coordination. If hetero-atomic coordination is preferred (chemically ordered arrangement, CO), the number of $AB$ pairs is maximized, so for $x<0.5$, \(P_{CO}(AA)=0\) and the other probabilities are: 
\[
P_{CO}(AB)=2x, \; P_{CO}(BB)=1-P_{CO}(AB),
\]while for $x \geq 0.5$ \(P_{CO}(BB)=0\) the relations are:  
\[
P_{CO}(AB)=2(1-x), \; P_{CO}(AA)=2x-1
\]
In the case of homo-atomic preference\textcolor{black}{ (H)}, the probabilities become  
\[
P_H(AA)=x, \; P_H(BB)=1-x,
\]
which corresponds to phase separation into $A$-rich and $B$-rich regions.
Let us now fix $A$ as the absorber and x the absorber concentration.  Since XAFS is chemically selective, it probes the conditional probability of finding a specific neighboring species given the absorber $A$. This is obtained by dividing the probability of the pair by the absorber concentration. In the case of a random distribution, the conditional probabilities are  
\[
P_R(A \vert A) = x, \qquad P_R(B \vert A) = 1-x.
\]
In the chemically ordered model, $B$ neighbors are maximized so for \(x \leq 0.5\), we obtain  
\[
P_{CO}(B \vert A) = 1, \qquad P_{CO}(A \vert A) = 0,
\]
while for \(x \geq 0.5\), the conditional probabilities are:  
\[
P_{CO}(B \vert A) = \frac{1}{x}-1, \qquad P_{CO}(A \vert A) = 2- \frac{1}{x}.
\]
In the BCTO EXAFS analysis, the multiplicity of Cu-Cu/Ta (and Ta-Ta/Cu) neighbors along the pseudo-cube is refined by constraining the total coordination number to 6 and weighted by the homo-atomic ($y = P(A \vert A)$) and hetero-atomic ($1-y = P(B\vert A)$) pair probabilities. The experimental results at the Cu and Ta edges are shown in Fig.\ref{fig:EXAFS_fit}d.
It is evident that, although the results carry a certain degree of uncertainty due to the complexity of the analysis and the difficulty of disentangling the different contributions, the experimental data suggest a clear tendency towards a hetero-atomic correlations and are consistently reproduced at both absorption edges. Such a preference for hetero-atomic correlation is not unexpected, as alternating Cu and Ta on the lattice reduces strain energy arising from the different shapes and sizes of the CuO$_6$ and TaO$_6$ octahedra.

\bibliography{mybib.bib}

@article{XANES_NbTa,
  title={Structural Analysis of Group V, VI, and VII Metal Compounds by XAFS},
  author={Asakura, Hiroyuki and Shishido, Tetsuya and Yamazoe, Seiji and Teramura, Kentaro and Tanaka, Tsunehiro},
  journal={J. Phys. Chem. C},
  volume={115},
  number={48},
  pages={23653-–23663},
  year={2011},
  publisher={APS},
  doi = {10.1021/jp2034104}
}

@article{MeneghiniPRL2009,
  title = {Nature of ``Disorder'' in the Ordered Double Perovskite \Uppercase{S}r2\Uppercase{F}e\Uppercase{M}o\Uppercase{O}6},
  author = {Meneghini, C. and Ray, Sugata and Liscio, F. and Bardelli, F. and Mobilio, S. and Sarma, D. D.},
  journal = {Phys. Rev. Lett.},
  volume = {103},
  issue = {4},
  pages = {046403},
  numpages = {4},
  year = {2009},
  month = {Jul},
  publisher = {American Physical Society},
  doi = {10.1103/PhysRevLett.103.046403},
  url = {https://link.aps.org/doi/10.1103/PhysRevLett.103.046403}
}

@article{Boyce1989,
title = {Local structure of pseudobinary semiconductor alloys: An x-ray absorption fine structure study},
journal = {Journal of Crystal Growth},
volume = {98},
number = {1},
pages = {37-43},
year = {1989},
issn = {0022-0248},
doi = {https://doi.org/10.1016/0022-0248(89)90183-8},
url = {https://www.sciencedirect.com/science/article/pii/0022024889901838},
author = {J.B. Boyce and J.C. Mikkelsen}
}

@article{SM,
    author = {Sagar Mahapatra et al.},
    title = {Exploring the random-single ground state in disrodered perovskites},
    journal = {Supplementary Material},
    year ={2025} 
}

@article{Coles01041978,
author = {B. R. Coles and B. V. B. Sarkissian and R. H. Taylor and},
title = {The role of finite magnetic clusters in Au-Fe alloys near the percolation concentration},
journal = {Philosophical Magazine B},
volume = {37},
number = {4},
pages = {489--498},
year = {1978},
publisher = {Taylor \& Francis}}

@article{3,
author = {Shirata ,Yutaka and Tanaka ,Hidekazu and Ono ,Toshio and Matsuo ,Akira and Kindo ,Koichi and Nakano ,Hiroki},
title = {Quantum Magnetization Plateau in Spin-1 Triangular-Lattice Antiferromagnet Ba3NiSb2O9},
journal = {Journal of the Physical Society of Japan},
volume = {80},
number = {9},
pages = {093702},
year = {2011},
doi = {10.1143/JPSJ.80.093702},

URL = { 
    
        https://doi.org/10.1143/JPSJ.80.093702
    
    

},
eprint = { 
    
        https://doi.org/10.1143/JPSJ.80.093702
    
    

}
}

@article{1,
doi = {10.1088/0953-8984/6/42/019},
url = {https://dx.doi.org/10.1088/0953-8984/6/42/019},
year = {1994},
month = {oct},
publisher = {},
volume = {6},
number = {42},
pages = {8891},
author = {A V Chubukov and  S Sachdev and  T Senthil},
title = {Large-S expansion for quantum antiferromagnets on a triangular lattice},
journal = {Journal of Physics: Condensed Matter}
}

@article{2,
  title = {Long-Range N\'eel Order in the Triangular Heisenberg Model},
  author = {Capriotti, Luca and Trumper, Adolfo E. and Sorella, Sandro},
  journal = {Phys. Rev. Lett.},
  volume = {82},
  issue = {19},
  pages = {3899--3902},
  numpages = {0},
  year = {1999},
  month = {May},
  publisher = {American Physical Society},
  doi = {10.1103/PhysRevLett.82.3899},
  url = {https://link.aps.org/doi/10.1103/PhysRevLett.82.3899}
}

@article{5,
title = {Resonating valence bonds: A new kind of insulator?},
journal = {Materials Research Bulletin},
volume = {8},
number = {2},
pages = {153-160},
year = {1973},
issn = {0025-5408},
doi = {https://doi.org/10.1016/0025-5408(73)90167-0},
url = {https://www.sciencedirect.com/science/article/pii/0025540873901670},
author = {P.W. Anderson}
}

@article{7_BCSO_Zhou,
  title = {Spin Liquid State in the \uppercase{S} = 1/2 Triangular Lattice \uppercase{B}a3\uppercase{C}u\uppercase{S}b2\uppercase{O}9},
  author = {Zhou, H. D. and Choi, E. S. and Li, G. and Balicas, L. and Wiebe, C. R. and Qiu, Y. and Copley, J. R. D. and Gardner, J. S.},
  journal = {Phys. Rev. Lett.},
  volume = {106},
  issue = {14},
  pages = {147204},
  numpages = {4},
  year = {2011},
  month = {Apr},
  publisher = {American Physical Society},
  doi = {10.1103/PhysRevLett.106.147204},
  url = {https://link.aps.org/doi/10.1103/PhysRevLett.106.147204}
}

@article{kimchi_2018,
  title = {Valence Bonds in Random Quantum Magnets: Theory and Application to \uppercase{Y}b\uppercase{M}g\uppercase{G}a\uppercase{O}4},
  author = {Kimchi, Itamar and Nahum, Adam and Senthil, T.},
  journal = {Phys. Rev. X},
  volume = {8},
  issue = {3},
  pages = {031028},
  numpages = {34},
  year = {2018},
  month = {Jul},
  publisher = {American Physical Society},
  doi = {10.1103/PhysRevX.8.031028},
  url = {https://link.aps.org/doi/10.1103/PhysRevX.8.031028}
}

@article{YbMg_2017,
  title={Disorder-induced mimicry of a spin liquid in \uppercase{Y}b\uppercase{M}g\uppercase{G}a\uppercase{O}4},
  author={Zhu, Zhenyue and Maksimov, PA and White, Steven R and Chernyshev, AL},
  journal={Physical review Letters},
  volume={119},
  number={15},
  pages={157201},
  year={2017},
  publisher={APS}
}

@article{18_SKundu_YCTO,
  title={Signatures of a Spin-1 2 Cooperative Paramagnet in the Diluted Triangular Lattice of \uppercase{Y}2\uppercase{C}u\uppercase{T}i\uppercase{O}6},
  author={Kundu, S and Hossain, Akmal and Das, Ranjan and Baenitz, M and Baker, Peter J and Orain, Jean-Christophe and Joshi, DC and Mathieu, Roland and Mahadevan, Priya and Pujari, Sumiran and others},
  journal={Physical review letters},
  volume={125},
  number={11},
  pages={117206},
  year={2020},
  publisher={APS}
}

@article{22_kimchi_nat_com_scaling,
  title={Scaling and data collapse from local moments in frustrated disordered quantum spin systems},
  author={Kimchi, Itamar and Sheckelton, John P and McQueen, Tyrel M and Lee, Patrick A},
  journal={Nature communications},
  volume={9},
  number={1},
  pages={4367},
  year={2018},
  publisher={Nature Publishing Group UK London}
}

@article{SrIr2O6,
  title = {Evidence for the random singlet phase in the honeycomb iridate \uppercase{S}r\uppercase{I}r2\uppercase{O}6},
author = {Song, Pengbo and Zhu, Kejia and Yang, Fan and Wei, Yuan and Zhang, Lu and Yang, Huaixin and Sheng, Xian-Lei and Qi, Yang and Ni, Jiamin and Li, Shiyan and Li, Yanchun and Cao, Guanghan and Meng, Zi Yang and Li, Wei and Shi, Youguo and Li, Shiliang},
  journal = {Phys. Rev. B},
  volume = {103},
  issue = {24},
  pages = {L241114},
  numpages = {7},
  year = {2021},
  month = {Jun},
  publisher = {American Physical Society},
  doi = {10.1103/PhysRevB.103.L241114},
  url = {https://link.aps.org/doi/10.1103/PhysRevB.103.L241114}
}

@article{Dasgupta_PhysRevB.22.1305,
  title = {Low-temperature properties of the random Heisenberg antiferromagnetic chain},
  author = {Dasgupta, Chandan and Ma, Shang Keng},
  journal = {Phys. Rev. B},
  volume = {22},
  issue = {3},
  pages = {1305--1319},
  numpages = {0},
  year = {1980},
  month = {Aug},
  publisher = {American Physical Society},
  doi = {10.1103/PhysRevB.22.1305},
  url = {https://link.aps.org/doi/10.1103/PhysRevB.22.1305}
}

@article{sana2024possible,
  title={Possible realization of a randomness-driven quantum disordered state in the \uppercase{S} = 1/2 antiferromagnet \uppercase{S}r3\uppercase{C}u\uppercase{T}a2\uppercase{O}9},
  author={Sana, B and Barik, M and Lee, S and Jena, U and Baenitz, M and Sichelschmidt, J and Luther, S and K{\"u}hne, H and Sethupathi, K and Rao, MS Ramachandra and others},
  journal={Physical Review B},
  volume={110},
  number={13},
  pages={134412},
  year={2024},
  publisher={APS}
}

@article{hossain2024evidence,
  title={Evidence of random spin-singlet state in the three-dimensional quantum spin liquid candidate \uppercase{S}r3\uppercase{C}u\uppercase{N}b2\uppercase{O}9},
  author={Hossain, SM and Rahaman, SS and Gujrati, H and Bhoi, Dilip and Matsuo, A and Kindo, K and Kumar, M and Majumder, M},
  journal={Physical Review B},
  volume={110},
  number={2},
  pages={L020406},
  year={2024},
  publisher={APS}
}

@article{LiZn2012,
  title={Possible valence-bond condensation in the frustrated cluster magnet \uppercase{L}i\uppercase{Z}n2\uppercase{M}o3\uppercase{O}8},
  author={Sheckelton, John P and Neilson, James R and Soltan, Daniel G and McQueen, Tyrel M},
  journal={Nature materials},
  volume={11},
  number={6},
  pages={493--496},
  year={2012},
  publisher={Nature Publishing Group UK London}
}

@article{YbZn2018,
  title={Spin-glass ground state in a triangular-lattice compound \uppercase{Y}b\uppercase{Z}n\uppercase{G}a\uppercase{O}4},
  author={Ma, Zhen and Wang, Jinghui and Dong, Zhao-Yang and Zhang, Jun and Li, Shichao and Zheng, Shu-Han and Yu, Yunjie and Wang, Wei and Che, Liqiang and Ran, Kejing and others},
  journal={Physical Review Letters},
  volume={120},
  number={8},
  pages={087201},
  year={2018},
  publisher={APS}
}

@article{BaY2010,
  title={Valence bond glass on an fcc lattice in the double perovskite \uppercase{B}a2\uppercase{Y}\uppercase{M}o\uppercase{O}6},
  author={de Vries, Mark A and Mclaughlin, AC and Bos, J-WG},
  journal={Physical review letters},
  volume={104},
  number={17},
  pages={177202},
  year={2010},
  publisher={APS}
}

@article{ma1979,
  title={Random antiferromagnetic chain},
  author={Ma, Shang Keng and Dasgupta, Chandan and Hu, Chin Kun},
  journal={Physical review letters},
  volume={43},
  number={19},
  pages={1434},
  year={1979},
  publisher={APS}
}

@article{meneghini2012estra,
  title={ESTRA-FitEXA: a software package for EXAFS data analysis},
  author={Meneghini, C and Bardelli, F and Mobilio, S},
  journal={Nuclear Instruments and Methods in Physics Research Section B: Beam Interactions with Materials and Atoms},
  volume={285},
  pages={153--157},
  year={2012},
  publisher={Elsevier}
}

@article{meneghini1994structure,
  title={Structure of $\alpha$-\uppercase{S}i1-x\uppercase{C}x: H alloys by wide-angle x-ray scattering: Detailed determination of first-and second-shell environment for Si and C atoms},
  author={Meneghini, C and Boscherini, F and Evangelisti, F and Mobilio, S},
  journal={Physical Review B},
  volume={50},
  number={16},
  pages={11535},
  year={1994},
  publisher={APS}
}

@article{Fisher1994,
  title={Random antiferromagnetic quantum spin chains},
  author={Fisher, Daniel S},
  journal={Physical Review B},
  volume={50},
  number={6},
  pages={3799},
  year={1994},
  publisher={APS}
}

\end{document}